\PassOptionsToPackage{square,numbers}{natbib} 
\documentclass[10pt]{article}
\usepackage[utf8]{inputenc}
\usepackage[T1]{fontenc}
\usepackage[english]{babel}
\usepackage{graphicx}
\usepackage{amsmath,amssymb,amsbsy,amsthm,color,natbib}
\usepackage{multicol}
\usepackage{color} 
\usepackage{natbib}
\usepackage[linktoc=all,colorlinks=true,citecolor=blue,linkcolor=blue]{hyperref}
\usepackage[paperwidth=18cm,paperheight=29.7cm,hmargin=2.5cm,vmargin=2.5cm]{geometry}
\usepackage{authblk}
\usepackage{comment}
\usepackage{cleveref}
\usepackage{subcaption}
\graphicspath{{figures/}}
\crefformat{equation}{#2{\color{red}#1}#3}
\Crefformat{equation}{#2{\itshape\color{red}Equation~#1}#3}
\crefformat{figure}{#2{\color{cyan}Fig.~#1}#3}
\crefformat{multifigures}{{#2{\color{cyan}Figs.~#1}#3}}
\Crefformat{figure}{#2{\color{cyan}Figure~#1}#3}
\Crefformat{multifigures}{{#2{\color{cyan}Figures~#1}#3}}
\crefformat{section}{#2{\color{cyan}Sect.~#1}#3}
\Crefformat{section}{#2{\color{cyan}Section~#1}#3}
\crefformat{chapter}{#2{\color{cyan}Chapt.~#1}#3}
\Crefformat{chapter}{#2{\color{cyan}Chapter~#1}#3}
\crefformat{table}{#2{\color{cyan}Table~#1}#3}
\Crefformat{table}{#2{\color{cyan}Table~#1}#3}

\title{Using Lagrangian descriptors to reveal the phase space structure of dynamical systems described by fractional differential equations: Application to the Duffing oscillator}

\author[1]{Dylan Theron}
\author[2]{Hadi Susanto}
\author[3]{Makrina Agaoglou}
\author[4]{Charalampos Skokos\thanks{: haris.skokos@uct.ac.za}}

\affil[1,4]{Nonlinear Dynamics and Chaos Group, Department of Mathematics and Applied Mathematics,\\ University of Cape Town, Rondebosch 7701, South Africa}
\affil[2]{Department of Mathematics, Khalifa University, PO Box 127788, Abu Dhabi, United Arab Emirates}
\affil[3]{Departamento de Matem\'atica Aplicada a la Ingenier\'ia Industrial, Escuela T\'ecnica Superior de Ingenieros Industriales,\\ Universidad Polit\'ecnica de Madrid, 28006 Madrid, Spain}

\begin{document}

\maketitle

\begin{abstract}
We showcase the utility of the Lagrangian descriptors method in qualitatively understanding the underlying dynamical behavior of dynamical systems governed by fractional-order differential equations. In particular, we use the Lagrangian descriptors method to study the phase space structure of the unforced and undamped Duffing oscillator when its time evolution is governed by fractional-order differential equations. In our study, we implement two types of fractional derivatives, namely the standard Gr\"unwald-Letnikov method, which is a finite difference approximation of the Riemann-Liouville fractional derivative, and a Gr\"unwald-Letnikov method with a correction term that approximates the Caputo fractional derivative. While there is no issue with forward-time integrations needed for the evaluation of Lagrangian descriptors, we discuss in detail ways to perform the non-trivial task of backward-time integrations and implement two methods for this purpose:  a `nonlocal implicit inverse' technique and a `time-reverse inverse' approach.  We analyze the differences in the Lagrangian descriptors results due to the two backward-time integration approaches,  discuss the physical significance of these differences, and eventually argue that the nonlocal implicit inverse implementation of the Gr\"unwald-Letnikov fractional derivative manages to reveal the phase space structure of fractional-order dynamical systems correctly.  
\end{abstract}

\section{Introduction}
\label{sec:intro}

Fractional calculus, a generalization of classical calculus, extends the concept of derivatives and integrals to non-integer (or fractional) orders. While conventional calculus, attributed to Newton and Leibniz, is well-suited for dealing with integer-order differentials, fractional calculus traces its origins back to Leibniz, who first introduced the concept in a letter to Guillaume de l'H\^opital in 1695, pondering the possibility of a half-order derivative \cite{ross1977development}.  Unlike classical derivatives, which depend solely on local properties, fractional derivatives account for the history of the function, making them well-suited to model complex systems with memory and hereditary properties, see, e.g., \cite{kilbas2006theory, tarasov2011fractional, herrmann2011fractional}. 

Over the years, fractional derivatives have been successfully used in studies of various physical systems. As an example, we mention the application of fractional derivatives in modeling the viscoelastic behavior of materials, for which traditional integer-order models often fail to capture their time-dependent behavior accurately \cite{frac2010,sasso2011application,frac2020}. Fractional derivative models have also been applied to describe the stress-strain relationship in polymers and elastomers, commonly used in various engineering applications, such as vibration dampers and seismic energy dissipation devices. This approach allows for a more accurate representation of the material's response under different loading conditions, including creep, relaxation, and cyclic loading tests \cite{bagley1983fractional,sasso2011application}. Furthermore, a lot of theoretical work has been done on the development of efficient numerical schemes for constructing and computing fractional derivatives and integrals, see, e.g.,~\cite{frac11,frac4,frac10,frac8,frac7,frac1,frac6,frac9,frac5,frac2,frac3}.

Despite the increase in utility, providing a physical or geometrical interpretation of fractional integration and differentiation has proven quite tricky. This lack of a suitable interpretation has been discussed in various conferences and meetings, some of which have been mentioned, e.g., in \cite{podlubny2001geometric}. Since those early conferences, multiple attempts have offered different explanations of fractional calculus. For example, numerous articles have tried to link fractional calculus and fractal geometry (see, e.g., \ \cite{nigmatullin1992fractional, yu1997fractional, moshrefi1998physical}). Additional examples of explanatory attempts include providing an interpretation of a fractional order Gr\"unwald-Letnikov differintegral by measuring the path and acceleration of a point in motion \cite{cioc2016physical}, or by first offering a straightforward geometric elucidation of fractional integrals, which is then used to introduce a physical understanding of the Riemann-Liouville fractional integration with respect to a dynamic time scale \cite{podlubny2001geometric}. Quoting Podlubny (2001), ``\dots it is difficult to speak about an acceptable geometric interpretation if one cannot see any picture there" \cite{podlubny2001geometric}. In this work, we offer an approach of qualitatively visualizing phase space structures of dynamical systems governed by fractional differential equations through implementing the Lagrangian descriptors method \cite{agaoglou2020lagrangian}. Our work is, therefore, expected to pave the way towards providing a physical and geometrical interpretation of fractional derivatives. 

The Lagrangian descriptors method is a tool that helps to identify phase space structures of systems governed by integer-order differential equations, including invariant manifolds \cite{daquin2022global}. These descriptors are scalar functions obtained by the time integration of a positive quantity, such as the modulus of velocity or acceleration, along particle trajectories over a finite time interval. By evaluating these quantities through the evolution of trajectories, both forward and backward in time, one can provide a comprehensive picture of the underlying dynamics in a computationally efficient way compared to traditional methods, like the computation of finite-time Lyapunov exponents \cite{mancho2013lagrangian}. The method of Lagrangian descriptors was first introduced in \cite{jMa09} and was initially inspired by efforts to explain the geometric patterns that control transport in geophysical flows. Since then, Lagrangian descriptors have found various applications in fields such as chemical reaction dynamics \cite{Craven,agaoglou2019lagrangian,agaoglou2020lagrangian,katsanikas2020detection,mAg21,skokos2,revuelta},  geophysical flows \cite{jMa09,cMe10,cMe12,geo2014,marine,curbelo1},  cardiovascular flows \cite{card}, biomedical flows \cite{Nar}, and billiard dynamics \cite{bil},  to name a few. In addition, Lagrangian descriptors have also been used as diagnostics of chaotic behavior \cite{daquin2022global,skokos3,skokos1,daquin2023a}. As a more general remark, we note that approaches based on the Lagrangian description of dynamical systems have been used to study phase space structures in diverse research fields ranging from plasma physics to oceanic flows \cite{plasmacoherent1,geocoherent1,plasmacoherent2,plasmacoherent3,geocohernet2}.  

In this paper, we apply the method of Lagrangian descriptors for the first time to systems governed by fractional-order differential equations to obtain their phase portraits, which would not be possible using traditional methods due to the nonlocal character of fractional-order differential equations. It is worth mentioning that this nonlocality creates difficulties in calculating the backward-in-time part of the Lagrangian descriptors, a topic that we also address in this work. 

The paper is organized as follows. In Sect.~\ref{NM}, we present the definitions of the various fractional time derivatives used in this paper, while in Sect.~\ref{Lds}, we present the Lagrangian descriptors method. Then, in  Sect.~\ref{R}, we use Lagrangian descriptors computations to visualize the phase space structure of the Duffing oscillator when its dynamics are described by fractional-order differential equations and present different approaches of performing backward time integrations. In addition, we discuss in detail which of these approaches manage to reveal the phase space structure of the system correctly. In Sect.~\ref{C}, we present our conclusions and discuss potential future work directions.

\section{Fractional derivatives}
\label{NM}

Unlike classical integer-order derivatives, fractional derivatives offer a more flexible framework that arises from how fractional calculus can be generalized from its classical counterpart. The concept of fractional differentiation is, therefore, not unique, and over the years, numerous definitions of fractional derivatives have been proposed \cite{teodoro2019review}. The most well-known definitions include the Riemann-Liouville fractional derivative, the Caputo derivative, and the Gr\"unwald-Letnikov derivative. Each of these definitions is based on distinct mathematical principles, yet they aim to extend the concept of differentiation to non-integer orders, e.g., \cite{li2015numerical}.

The Riemann-Liouville fractional derivative is among the oldest and most widely used definitions. It is based on an integral operator and offers a natural extension of integer-order differentiation. Let $C^n(I,\mathbb{R})$ denote the space of $n$-times continuously differentiable single variable real-valued 
functions on the real 
interval $I$. We also let $\alpha\in(0,\infty)\setminus \mathbb{N}$ and the function $g\in C^n([a,b],\mathbb{R})$. The (left-sided) Riemann-Liouville fractional-order derivative of order $\alpha$ of the function $g$ at point $t$  is given by
\begin{equation}
\label{eq:RL_D}
    ^{RL}D_{a+}^{\alpha}g(t)=\frac{1}{\Gamma(n-\alpha)}\frac{d^n}{dt^n}\int_
a^t(t-s)^{n-\alpha-1}g(s)ds,
\end{equation}
where $n=\lfloor\alpha\rfloor+1$ and $\Gamma(x)$ is the usual Gamma function. However, this definition of fractional derivatives presents challenges in practical applications for the following reasons.

In initial value problems (IVPs) with classical (integer-order) differential equations, initial conditions (ICs) typically specify the function's values and integer-order derivatives at a given point. However, IVPs with non-integer order Riemann-Liouville derivatives often require ICs involving integrals or fractional integrals of the function 
\cite{kilbas2006theory,podlubny1998fractional,diethelm2010analysis}. As a result, standard ICs (i.e., specifying the value of the function at a single point) may not be sufficient, and one might need to specify additional conditions based on 
past behavior. 
This requirement adds complexity to solving fractional IVPs with the Riemann-Liouville definition. In that scenario, interpreting and applying this type of fractional derivatives in real-world scenarios where initial values are typically simple, such as initial positions or velocities in mechanical systems, becomes challenging.

Caputo modified the classical Riemann-Liouville fractional derivative to address the practical issues related to the ICs. The primary motivation behind Caputo's formulation was to make fractional derivatives more suitable for modeling physical processes, i.e., IVPs, particularly in applications where classical ICs (such as those used in integer-order differential equations) are commonly used. The (left-sided) Caputo fractional-order derivative of order $\alpha$ of a  function $g$ is given by
\begin{equation}
\label{eq:Caputo}
^CD_{a+}^{\alpha}g(t)=\frac{1}{\Gamma(n-\alpha)}\int_
a^t(t-s)^{n-\alpha-1}\frac{d^ng(s)}{ds^n}ds.
\end{equation}
The Caputo derivative can be written in terms of the Riemann-Liouville derivative as follows
\begin{equation}
^{RL}D^\alpha g(t)={}^CD^\alpha g(t) +\sum_{k=0}^{n-1}\frac{g^{(k)}(0^+)}{\Gamma(k-\alpha+1)}t^{k-\alpha}.
\label{RL-C}
\end{equation}
These two fractional derivatives coincide when the ICs are zero (homogeneous), i.e., $g^{(k)}(0^+)=0$, $k=0,\dots,n-1$, see e.g., \cite{scherer2011grunwald}. 
While the Caputo derivative is often preferred in modeling physical systems, the Riemann-Liouville derivative is favored for its historical significance and analytical tractability. 

Gr\"unwald and Letnikov generalized the notion of finite differences to non-integer orders. They worked on extending the concept of differentiation and integration to fractional orders using finite differences, naturally leading to a discretized form of fractional derivatives. The Gr\"unwald-Letnikov fractional derivative of order $\alpha$ of function $g$ is given by
\begin{equation}
^{GL}D_{a+}^{\alpha}g(t)=\lim_{h\rightarrow 0}h^{-\alpha}\sum_{r=0}^{\left[(t-a)/h\right]}(-1)^r\left(\begin{array}{c}
\alpha \\
r\end{array}
\right)g(t-rh),
\label{grunwald}
\end{equation}
where the generalized binomial coefficient is
\[
\binom{\alpha}{k} = \frac{\alpha (\alpha - 1) (\alpha - 2) \dots (\alpha - k + 1)}{k!}=\frac{\Gamma(\alpha+1)}{\Gamma(k+1)\Gamma(\alpha-k+1)}.
\]
It is important to note that if $\alpha\in(0,\infty)\setminus \mathbb{N}$ and $g:(0,\infty)\rightarrow \mathbb{R}$  
is a function of class $C^n$, then \cite{kilbas2006theory,podlubny1998fractional}
\begin{equation}\label{rel.Caputo.RL}
{}^{RL}D^\alpha g(t)={}^{GL}D^\alpha g(t). 
\end{equation}
The Gr\"unwald-Letnikov derivative \eqref{grunwald}, therefore, provides a framework for the numerical computation of the Riemann-Liouville fractional derivative. However, since the Riemann-Liouville derivative was formally introduced slightly later, Gr\"unwald and Letnikov did not set their work explicitly as its discretization. Using  \eqref{RL-C}, we can also obtain the Gr\"unwald-Letnikov representation for the Caputo derivative \cite{scherer2011grunwald}. We consider the Riemann-Liouville and Caputo derivatives using their Gr\"unwald-Letnikov representations in this work.

\section{Lagrangian descriptors}
\label{Lds}

The Lagrangian descriptors method involves summing, for any IC of a dynamical system, the values of a positive scalar function that depends on phase space variables along its trajectory, both forward and backward in time. This process is applied to a grid of ICs on a specific phase space slice to uncover the underlying dynamical structure. Furthermore, in the case of dynamical systems described by integer-order differential equations or of discrete-time systems, the resulting scalar field of Lagrangian descriptors highlights invariant stable and unstable manifolds, which appear as `singular features' where the descriptor values change abruptly. Forward trajectory integration identifies stable manifolds, while backward trajectory evolution identifies unstable manifolds. In this paper, we use the so-called $p$-`norm' definition of the Lagrangian descriptor \cite{lopesino2017lagrangian}
\begin{equation}
    M_p(\mathbf{x}_0,t_0,\tau)=\int_{t_0-\tau}^{t_0+\tau}\sum_{i=1}^{N}\big|f_i(\mathbf{x},t)\big|^pdt,
    \label{p-norm LD definition}
\end{equation}
where $f_i(\mathbf{x},t)$ is the $i^{\text{th}}$ component of a vector field, $t_0$ is the time at which we start the evolution of the studied trajectory (in this study we set $t_0=0$), $\tau$ is the integration time, and $\mathbf{x}_0$ is the IC of the considered orbit. We select parameter $p$ to allow for the greatest discontinuity of the gradient of the Lagrangian descriptor values at the manifold, thus permitting us to extract the normally hyperbolic invariant manifolds (NHIMs) \cite{eldering2013normally} from the Lagrangian descriptors scalar field. It has been shown that this occurs at $p=0.5$ for a wide variety of dynamical systems \cite{demian2017detection, katsanikas2020phase}. We note that, since $p<1$, the integrand in \eqref{p-norm LD definition} is not an actual norm, and for this reason, we use inverted commas over the word norm in the definition.

Furthermore, the Lagrangian descriptors method can be split, respectively, into its forward and backward components as
\begin{equation}    M_p^f(\mathbf{x}_0,t_0,\tau)=\int_{t_0}^{t_0+\tau}\sum_{i=1}^{N}\big|f_i(\mathbf{x},t)\big|^pdt,
    \label{forward p-norm LD definition}
\end{equation}
where $f$ indicates the forward method and 
\begin{equation}
    M_p^b(\mathbf{x}_0,t_0,\tau)=\int_{t_0-\tau}^{t_0}\sum_{i=1}^{N}\big|f_i(\mathbf{x},t)\big|^pdt,
    \label{backward p-norm LD definition}
\end{equation}
where $b$ specifies that this is the backward Lagrangian descriptors method. This formulation has been used for dynamical systems governed by ordinary differential equations. In our case, our model is governed by fractional-order differential equations that we solve as a discrete-time dynamical system. For this reason, we implement the $p$-`norm' discrete Lagrangian descriptors method in the following form (see,  e.g., \cite{lopesino2015lagrangian,mAg21,daquin2022global}) 
\begin{equation}
       MD_{p} (\mathbf{x_{0}},N)=h^{1-p}\sum_{j=-N}^{N-1}\sum_{i=1}^{k}|x_{j+1}^{i}-x_{j}^{i}|^{p},
    \label{p-norm discrete LD definition}
\end{equation}
where $x_j^i$ represents the $i^\text{th}$ component of the state vector $\mathbf{x}$ at time $t_j$ and $h=t_{j+1}-t_j$ is the discretization time step. The summation can also be split between the forward evolution of the orbit with IC $\mathbf{x_{0}}$
\begin{equation}
    MD_{p}^{+}(\mathbf{x_{0}},N) =h^{1-p}\sum_{j=0}^{N-1}\sum_{i=1}^{k}|x_{j+1}^{i}-x_{j}^{i}|^{p},
\end{equation}
and the backward evolution of the orbit $\mathbf{x_{0}}$
\begin{equation}
    MD_{p}^{-}(\mathbf{x_{0}},N)=h^{1-p}\sum_{j=-N}^{-1}\sum_{i=1}^{k}|x_{j+1}^{i}-x_{j}^{i}|^{p}.
\end{equation}

\section{Numerical results}
\label{R}

To illustrate the idea of exploiting the Lagrangian descriptors method for revealing the phase portrait of dynamical systems with fractional derivatives, we  consider the unforced and undamped Duffing oscillator with fractional-order derivatives
\begin{equation}
    \begin{split}
        D^{\alpha}x(t)&=y(t), \\
        D^{\alpha}y(t)&=x(t)-x(t)^3,
    \end{split}
    \label{fractional differential equation duffing oscillator}
\end{equation}
where the value of $\alpha$ in the differential operator $D^{\alpha}$ is the fractional order of the equations of motion. When $\alpha=1$, we obtain the classical Duffing oscillator originally described in \cite{duffing1918erzwungene}. When $\alpha$ in \eqref{fractional differential equation duffing oscillator} is a non-integer, we obtain a fractional Duffing oscillator that generalizes the concept of acceleration, where the system's present motion is influenced by both its current state and history. The fractional derivative can also be viewed as an interpolation between integer-order derivatives. For example, for $0.5<\alpha<1$, 
the fractional derivative can be viewed to lie between the first derivative (velocity) and the second derivative (acceleration), capturing intermediate dynamical effects. Duffing oscillator systems with damping and drive have been studied in, e.g., \cite{li2015nonlinear,ilhan2022interesting,li2022cluster}, focusing on systems' chaotic behavior, their bifurcation analysis, and the observed multistability due to the fractionality. Another way of introducing fractionality in the Duffing oscillator is by applying it to its damping term rather than the acceleration, reflecting the need to model systems with complex energy dissipation, see, e.g., \cite{jimenez2013fractional,xu2013responses,jimenez2015characterizing,torkzadeh2021numerical,rysak2022damping,hamaizia2024rich}. In these cases, the dissipation of energy is gradual and influenced by past states of the system that can lead to rich dynamics, including slow decay rates. Nevertheless, trying to demonstrate the applicability and usefulness of the Lagrangian descriptors method to dynamical systems of fractional derivatives, we consider in our study the simplest possible version of the Duffing oscillator \eqref{fractional differential equation duffing oscillator} by neglecting the possible effects of damping and external forcing.

\subsection{Phase space structures}
\label{sec:R_color}

\begin{figure}[tb!]
    \centering
    \begin{subfigure}{0.75\textwidth}
\text{\qquad\qquad$\tau=5$\qquad\qquad\qquad\qquad$\tau=10$\qquad\qquad\qquad$\tau=20$}\par\smallskip
        \includegraphics[width=\textwidth]{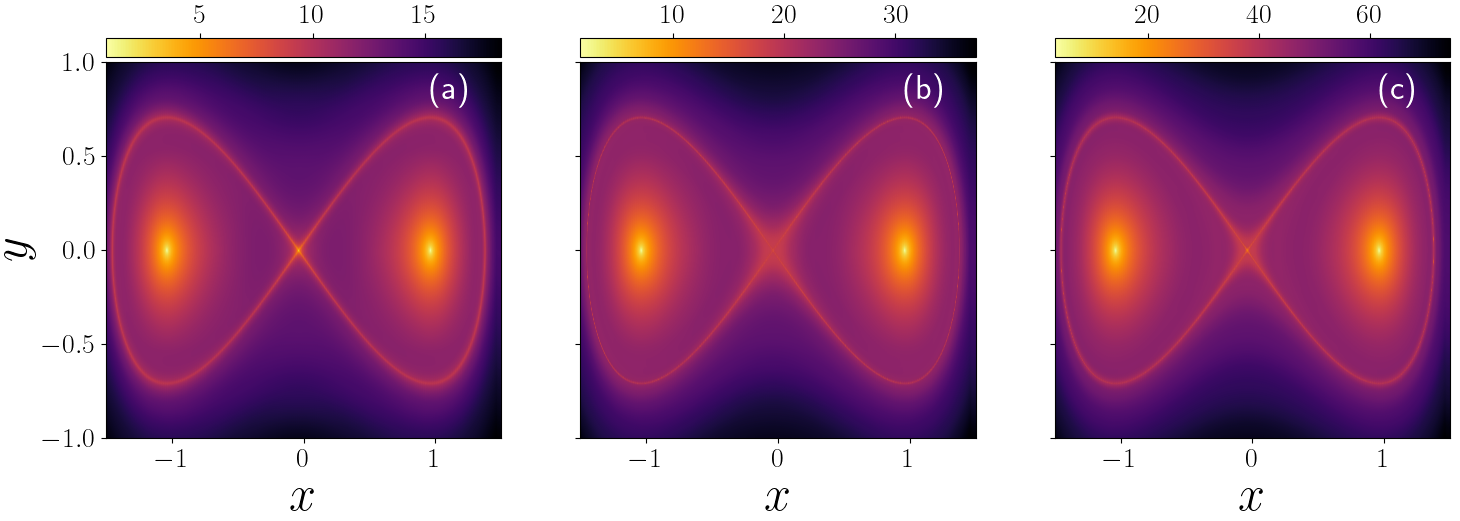}    
    \end{subfigure}
    \caption{Initial conditions colored according to their $p$-`norm' Lagrangian descriptor value \eqref{p-norm LD definition}, by using  the color bar above each panel, of the classical Duffing oscillator, i.e., system \eqref{fractional differential equation duffing oscillator} for $\alpha=1$. The three panels are constructed for a grid of $1000\times1000$ equidistant ICs over the intervals $x\in[-1.5,1.5]$ and $y\in[-1,1]$, and show results for an integration time of (a) $\tau=5$, (b) $\tau=10$, and (c) $\tau=20$.}
    \label{LDs method for Duffing Oscillator with ODEs}
    \label[multifigures]{LDs method for Duffing Oscillator with ODEs multiple}
\end{figure}

The Lagrangian descriptors method has already been applied to the classical Duffing oscillator 
\eqref{fractional differential equation duffing oscillator} with $\alpha=1$ \cite{agaoglou2020lagrangian}, and variations of this dynamical system, e.g., the Duffing oscillator described by stochastic equations of motion \cite{balibrea2016lagrangian}. The method provided a scalar field representation highlighting key geometric structures in the system's phase space, such as saddle points, attractors, repellers, and NHIMs. By integrating  trajectories over a finite time interval, the method managed to visualize these critical structures without requiring exhaustive long-time computations. In \cref{LDs method for Duffing Oscillator with ODEs} we present results obtained for the classical Duffing oscillator by applying the $p$-`norm' definition of the Lagrangian descriptors method \eqref{p-norm LD definition} to a grid of $1000\times1000$ equidistant ICs over the intervals $x\in[-1.5,1.5]$ and $y\in[-1,1]$ with an integration time of $\tau=5$ [\cref{LDs method for Duffing Oscillator with ODEs}(a)], $\tau=10$ [\cref{LDs method for Duffing Oscillator with ODEs}(b)], and  $\tau=20$ [\cref{LDs method for Duffing Oscillator with ODEs}(c)]. The color bars above each panel correspond to the magnitude of the Lagrangian descriptor values in that plot. From the results of \cref{LDs method for Duffing Oscillator with ODEs} we easily verify the existence of  three fixed points, namely two stable points at $(x,y)=(\pm1,0)$ and one unstable point at $(x,y)=(0,0)$. We also see that the stable and unstable manifolds of the origin are connected, and are  creating an ``infinity shaped" structure. In addition, we observe the existence of  two regions with low Lagrangian descriptor values near the fixed points at $(x,y)=(-1,0)$ and $(x,y)=(1,0)$. The NHIMs are seen as curves within the Lagrangian descriptor scalar field, at which Lagrangian descriptor values exhibit a sudden change (i.e., points that have a sharp change in gradient \cite{katsanikas2020detection}). The reason that we see approximately the same shape of NHIMs in all panels of \cref{LDs method for Duffing Oscillator with ODEs} is due to the invariant nature of the NHIMs (see, e.g., \cite{eldering2013normally}). Additionally, the magnitude of the Lagrangian descriptor values increase from \cref{LDs method for Duffing Oscillator with ODEs}(a) ($\tau=5$) to \cref{LDs method for Duffing Oscillator with ODEs}(c) ($\tau=20$) because of the longer integration time, which leads to the accumulation of more positive values in the computation of the descriptors. We will use the results of \cref{LDs method for Duffing Oscillator with ODEs}, obtained for the integer derivative $\alpha=1$ case of \eqref{fractional differential equation duffing oscillator}, as a reference case to examine the changes in a system's phase space structures caused by different orders of the used fractional derivatives, i.e., for $0<\alpha<1$. 

The challenge in obtaining the Lagrangian descriptors for the case of the fractional derivatives is calculating the part of the method for which we need to integrate the equations of motion \eqref{fractional differential equation duffing oscillator} backward in time [see Eq.~\eqref{backward p-norm LD definition}]. The integration backward in time (or the inverse time integration) corresponds to solving the differential equations of motion by stepping backward along the time axis, effectively reversing the time evolution of a system. This approach is beneficial in specific problems, such as reconstructing past states from present data or sensitivity analysis, where understanding how a system evolved to its current state is critical. However, backward integration is often more challenging than forward integration due to numerical stability issues. Many physical systems governed by dissipative processes are inherently irreversible, meaning minor numerical errors during backward integration can grow exponentially, leading to inaccurate or non-physical solutions. 

To perform inverse time integration for dynamical systems described by integer-order derivatives, we can reverse the direction of the time variable and integrate the dynamical system by using the same numerical methods but with negative time steps. In the same vein, we obtain from \eqref{fractional differential equation duffing oscillator} the Duffing oscillator's backward dynamics
\begin{equation}
    \begin{split}
        D^{\alpha}x(t)&=-y(t), \\
        D^{\alpha}y(t)&=-x(t)+x(t)^3,
    \end{split}
    \label{time-reversing}
\end{equation}
by the transformation $t\to(-t)$, which can be evolved using the same one-step method as used for \eqref{fractional differential equation duffing oscillator} (e.g., the Gr\"unwald-Letnikov or Caputo derivative).
However, for fractional-order derivatives \eqref{time-reversing} no longer provides the backward trajectory of the system \cite{campos2015time}, 
because fractional-order derivatives inherently incorporate memory effects, i.e., the system's future states depend on the current and all previous states. This memory property adds complexity when attempting to reverse time. 

To determine the initial state $\mathbf{x}=\mathbf{x_{0}}$ at time $t=0$ from a given future state $\mathbf{x_{N}}$ at $t=t_N$, we must solve equations that depend on all unknown past values of the state vector $\mathbf{x}$ between $t=0$ and $t=t_N$. Therefore, the backward time integration method is implicit for the Gr\"unwald-Letnikov derivative \eqref{grunwald}. In this context, we must simultaneously solve a set of algebraic nonlinear equations for all time steps, and to do that, we employ a Newton-Raphson method. In the following, we refer to this approach as the `nonlocal implicit inverse' method. We note that when we implemented this approach, we did not obtain numerical convergence of the algorithm for some ICs. 
This problem of non-convergence became even more significant when we implemented the Caputo version of the fractional derivative \eqref{eq:Caputo}, with the `nonlocal implicit inverse' method. 
Because of that, for the sake of comparison, we use \eqref{time-reversing} for the Caputo derivatives with fractional $\alpha$. 
In the following, we call this latter approach, i.e., the transformation $t\to(-t)$, 
the `time-reversing inverse' method. 

\begin{figure}[tbhp!]
    \centering
    \begin{subfigure}{1\textwidth}
    \text{\qquad\qquad\qquad~~~$\tau=5$\qquad\qquad\qquad$\tau=5$\qquad\qquad\qquad$\tau=10$\qquad\qquad\qquad$\tau=20$}\par\smallskip
        \includegraphics[width=\textwidth]{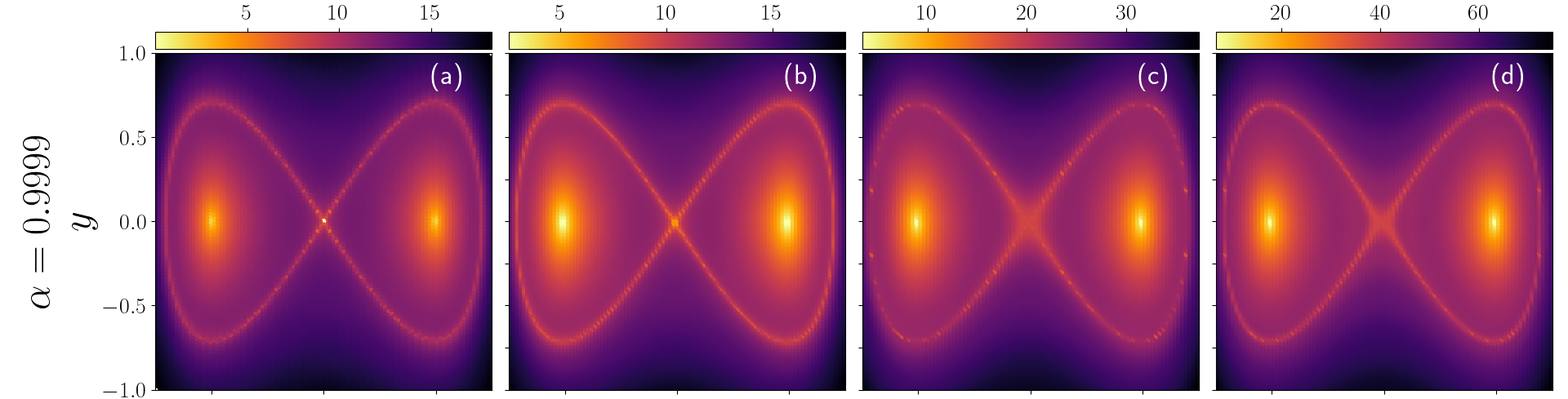}    
    \end{subfigure}\hfill
    \begin{subfigure}{1\textwidth}
        \includegraphics[width=\textwidth]{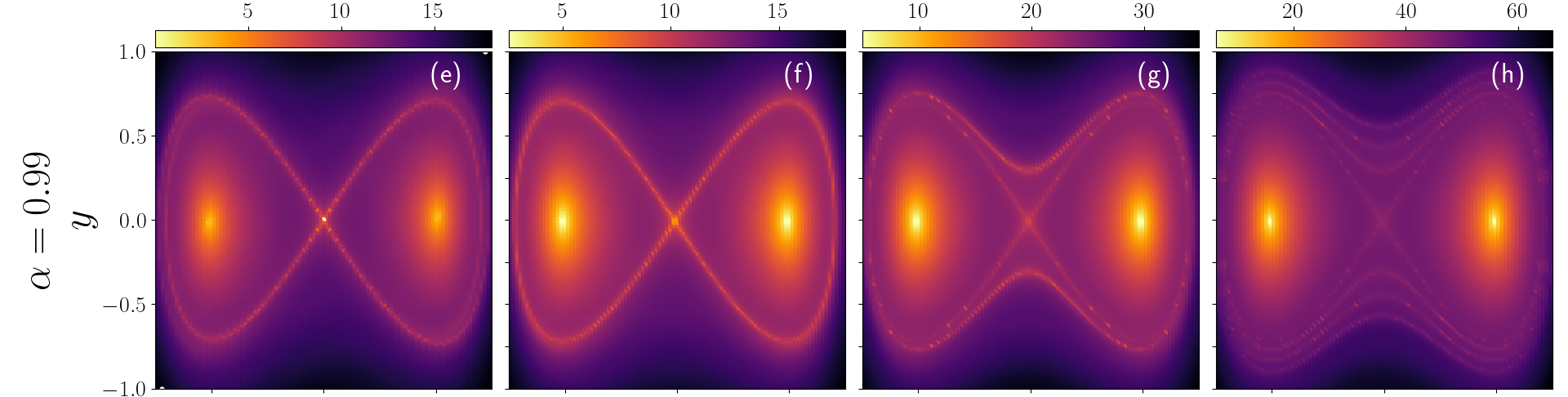}    
    \end{subfigure}\hfill
    \begin{subfigure}{1\textwidth}
        \includegraphics[width=\textwidth]{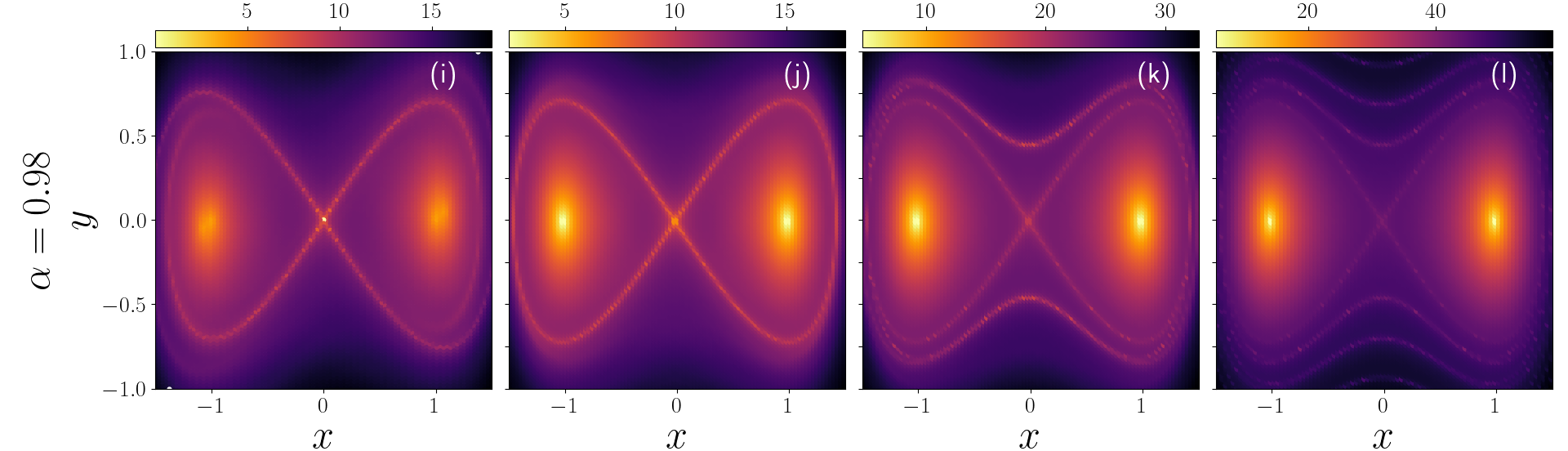}    
    \end{subfigure}\hfill
    \caption{The phase space of the Duffing oscillator governed by the fractional differential equations \eqref{fractional differential equation duffing oscillator} for different fractional orders $\alpha$ (indicated on the left of each row) and integration times $\tau$ (indicated on the top of each column),  where each IC is colored according to its Lagrangian descriptor value (using the color bar above each panel). Each plot is created using $100\times100$ equidistant ICs over the intervals $x\in[-1.5,1.5]$ and $y\in[-1,1]$. The first column of results [panels (a), (e), and (i)] reveals the phase space structure using the  Gr\"unwald-Letnikov method \eqref{grunwald} following the `nonlocal implicit inverse' approach (see text for details). The remaining panels in the second, third, and fourth columns correspond to results obtained by the `time-reversing inverse' method using the Caputo derivative. The first and second columns [panels (a), (b), (e), (f), (i), and (j)] show phase space features for integration time $\tau=5$, while the results of the third  [plots (c), (g), and (k)] and the fourth column [panels (d), (h), and (l)] were obtained for $\tau=10$ and $\tau=20$ respectively. Each row corresponds to the Duffing oscillator \eqref{fractional differential equation duffing oscillator} for different fractional orders: $\alpha=0.9999$ [top row, panels (a)--(d)], $\alpha=0.99$ [middle row, panels (e)--(h)], and $\alpha=0.98$ [bottom row, panels (i)--(l)].}
    \label{LDs for Duffing Oscillator with fractional differential equations of order close to one}
    \label[multifigures]{LDs for Duffing Oscillator with fractional differential equations of order close to one multiple}
\end{figure}

We now apply the $p$-`norm' Lagrangian descriptors method \eqref{p-norm LD definition} to the Duffing oscillator described by fractional-order differential equations (\cref{fractional differential equation duffing oscillator}) to examine the system's phase space formations. The obtained results are presented in \cref{LDs for Duffing Oscillator with fractional differential equations of order close to one}, for various integration times ($\tau=5,~10,$ and $20$), and for fractional orders of the dynamical system which are close to the classical Duffing oscillator ($\alpha=0.9999,~0.99,$ and $0.98$). In order to facilitate the direct comparison with the results obtained for $\alpha=1$ (\cref{LDs method for Duffing Oscillator with ODEs}) all panels of  \cref{LDs for Duffing Oscillator with fractional differential equations of order close to one} were created using a grid of $100\times100$ equidistant ICs over the same interval we considered in \cref{LDs method for Duffing Oscillator with ODEs}. We emphasize that we used a sparsely populated grid ($100^2$ grid points in  \cref{LDs for Duffing Oscillator with fractional differential equations of order close to one}  instead of $1000^2$ grid points in \cref{LDs method for Duffing Oscillator with ODEs}) due to the significantly larger CPU times required to integrate the fractional differential equations. As has already been mentioned, we considered two definitions of fractional derivatives: a) the Gr\"unwald-Letnikov method \eqref{grunwald} which approximates the Riemann-Liouville-type derivatives with the `nonlocal implicit inverse' approach, and b) the Caputo derivative \eqref{eq:Caputo} with the `time-reversing inverse' setting. 

The results obtained by the implementation of the  `nonlocal implicit inverse' method using the Gr\"unwald-Letnikov fractional derivative are shown in the first column of  \cref{LDs for Duffing Oscillator with fractional differential equations of order close to one} [panels (a), (e), and (i)] for an integration time of  $\tau=5$. The remaining plots in columns two, three, and four of \cref{LDs for Duffing Oscillator with fractional differential equations of order close to one} show the Lagrangian descriptor values obtained by the application of the `time-reversing inverse' method using the Caputo derivative. In particular, the results of column two [\cref{LDs for Duffing Oscillator with fractional differential equations of order close to one multiple}(b), (f), and (j)], three [\cref{LDs for Duffing Oscillator with fractional differential equations of order close to one multiple}(c), (g), and (k)], and four [\cref{LDs for Duffing Oscillator with fractional differential equations of order close to one multiple}(d), (h), and (l)] respectively correspond to integration times $\tau=5$, $\tau=10$, and  $\tau=20$. We note that we only used an integration time of $\tau=5$ for obtaining results by the `nonlocal implicit inverse' method [\cref{LDs for Duffing Oscillator with fractional differential equations of order close to one multiple}(a), (e), and (i)] as this approach requires significantly more CPU time compared to the `time-reverse inverse' [columns two, three, and four of \cref{LDs for Duffing Oscillator with fractional differential equations of order close to one}]. 

Each row of plots in \cref{LDs for Duffing Oscillator with fractional differential equations of order close to one} showcases the phase space formations of the Duffing oscillator \eqref{fractional differential equation duffing oscillator} for different fractional orders $\alpha$, namely $\alpha=0.9999$ [top row; \cref{LDs for Duffing Oscillator with fractional differential equations of order close to one multiple}(a)--(d)], $\alpha=0.99$ [middle row; \cref{LDs for Duffing Oscillator with fractional differential equations of order close to one multiple}(e)--(h)], and $\alpha=0.98$ [bottom row; \cref{LDs for Duffing Oscillator with fractional differential equations of order close to one multiple}(i)--(l)]. In all panels of \cref{LDs for Duffing Oscillator with fractional differential equations of order close to one multiple}(a)--(d), we see similar infinity-shaped phase space structures. This similarity indicates that for $\alpha=0.9999$ (a value which is very close to the integer-order case $\alpha=1$), the formations revealed by the Lagrangian descriptors method are approximately invariant both to the chosen integration time $\tau$ and the fractional derivative method used. Additionally, we see that the almost identical shapes of the phase space patterns in \cref{LDs for Duffing Oscillator with fractional differential equations of order close to one multiple}(a)--(d) are similar to those seen in \cref{LDs method for Duffing Oscillator with ODEs}. We also note that the ranges of Lagrangian descriptor values (indicated at the color bars above each panel of \cref{LDs for Duffing Oscillator with fractional differential equations of order close to one} are approximately equivalent to those seen in the panels of \cref{LDs method for Duffing Oscillator with ODEs} with the same integration time $\tau$.

Slightly reducing the fractional order of the dynamical system to $\alpha=0.99$ for the `nonlocal implicit inverse' method [\cref{LDs for Duffing Oscillator with fractional differential equations of order close to one}(e)] results in an infinity-shaped structure which is no longer continuous and connected, as in \cref{LDs for Duffing Oscillator with fractional differential equations of order close to one}(a), but exhibits some breaks at both edges of the $x$--axis. We also note that the phase portrait of \cref{LDs for Duffing Oscillator with fractional differential equations of order close to one}(e) has a point symmetry around the origin. In the plots obtained by the `time-reversing inverse' method [\cref{LDs for Duffing Oscillator with fractional differential equations of order close to one multiple}(f)--(h)], we see that the structures revealed by the Lagrangian descriptors method now show a clear dependency on $\tau$. The phase space formations for $\tau=5$ [\cref{LDs for Duffing Oscillator with fractional differential equations of order close to one}(f)] look very similar to those seen for  $\alpha=0.9999$ for the same integration time [\cref{LDs for Duffing Oscillator with fractional differential equations of order close to one}(b)]. By increasing the value of the integration time to $\tau=10$ [\cref{LDs for Duffing Oscillator with fractional differential equations of order close to one}(g)] we observe that, apart from the infinity-shaped feature, an additional curve outside of this formation is present, while the further increase to $\tau=0.99$ [\cref{LDs for Duffing Oscillator with fractional differential equations of order close to one}(h)] results in the appearance of many more such curves. It is worth noting that the structures created by the `time-reversing inverse' approach in \cref{LDs for Duffing Oscillator with fractional differential equations of order close to one multiple}(f)--(h) demonstrate reflective symmetry with respect to the horizontal and vertical axes passing through the origin $(x,y)=(0,0)$. The further decrease of the fractional derivative order to $\alpha = 0.98$ leads to an infinity-like structure with a more pronounced break when the `nonlocal implicit inverse' approach is used [\cref{LDs for Duffing Oscillator with fractional differential equations of order close to one}(i)]. In the results obtained by the `time-reversing inverse' methods [\cref{LDs for Duffing Oscillator with fractional differential equations of order close to one multiple}(j)--(l)] we witness similar geometric patterns to those presented in \cref{LDs for Duffing Oscillator with fractional differential equations of order close to one multiple}(f)--(h) for $\alpha = 0.99$, but the distances between the additional curves for $\tau=10$ [\cref{LDs for Duffing Oscillator with fractional differential equations of order close to one}(k)] and $\tau=20$ [\cref{LDs for Duffing Oscillator with fractional differential equations of order close to one}(l)] are increased with respect to what was observed in  \cref{LDs for Duffing Oscillator with fractional differential equations of order close to one multiple}(g) and (h).

\begin{figure}[tb!]
    \centering
    \begin{subfigure}{1\textwidth}
\text{\qquad\qquad\qquad~~~$\tau=5$\qquad\qquad\qquad$\tau=5$\qquad\qquad\qquad$\tau=10$\qquad\qquad\qquad$\tau=20$}\par\smallskip
        \includegraphics[width=\textwidth]{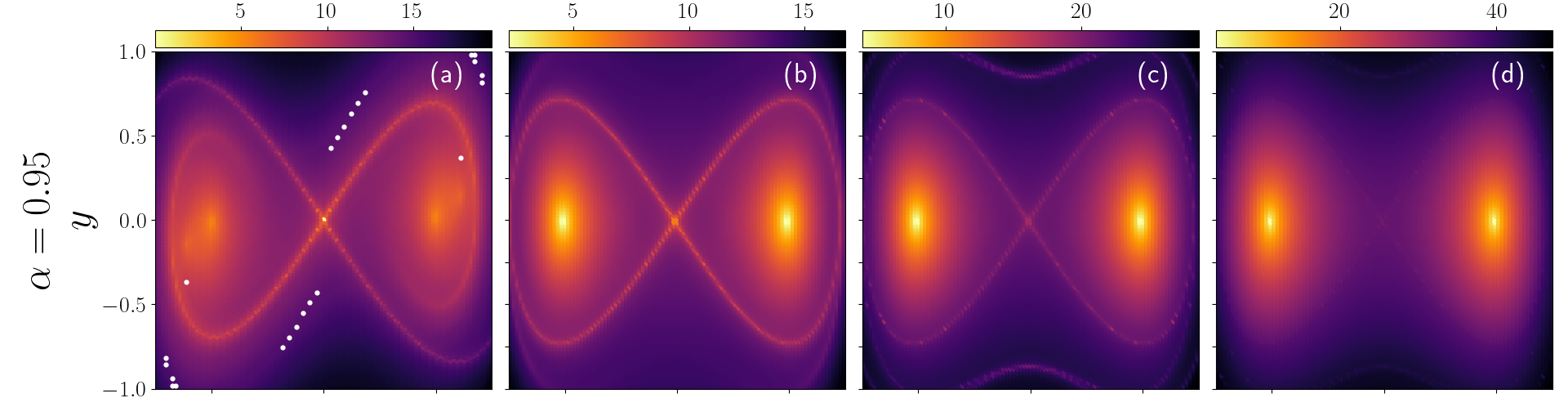}    
    \end{subfigure}\hfill
    \begin{subfigure}{1\textwidth}
        \includegraphics[width=\textwidth]{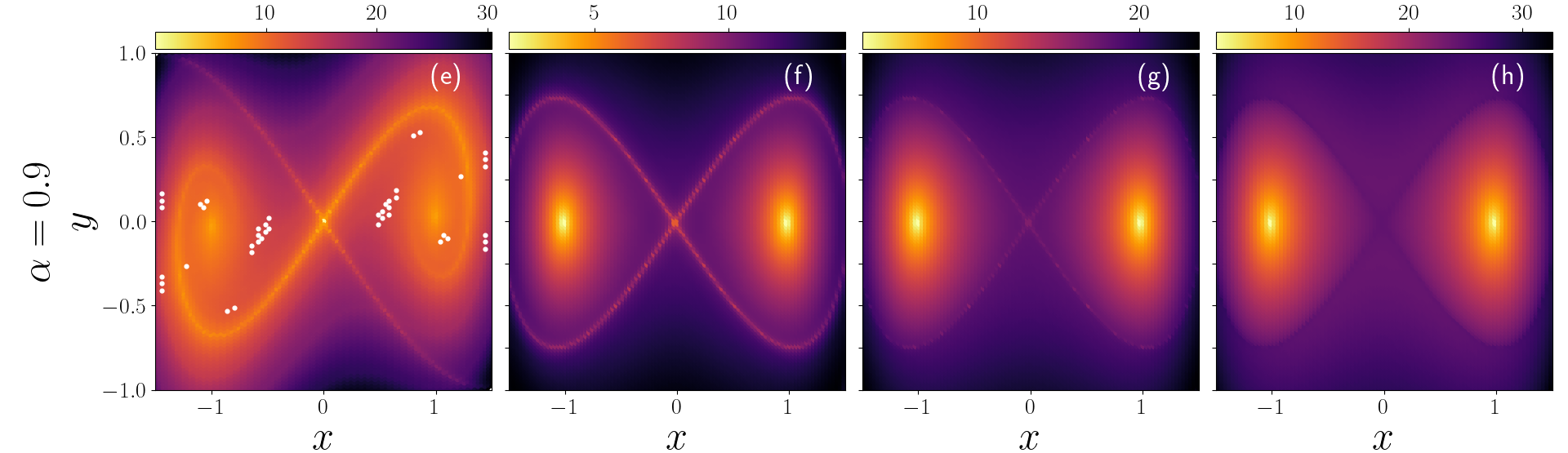}    
    \end{subfigure}\hfill
    \caption{Plots similar to those seen in \cref{LDs for Duffing Oscillator with fractional differential equations of order close to one}, but for $\alpha=0.95$ [first row, panels (a)--(d)] and $\alpha=0.9$ [second row, panels (e)--(h)]. White-colored points in panels (a) and (e) correspond to ICs whose backward time evolution was not computed due to the numerical instabilities of the implemented algorithm.}
    \label{LDs for Duffing Oscillator with fractional differential equations of order 0.95 and 0.9}
    \label[multifigures]{LDs for Duffing Oscillator with fractional differential equations of order 0.95 and 0.9 multiple}
\end{figure}

From the results of \cref{LDs for Duffing Oscillator with fractional differential equations of order close to one}, we see that  $\alpha \approx 1$ produces phase space structures similar to those observed for the integer order ($\alpha=1$)  Duffing oscillator in  \cref{LDs method for Duffing Oscillator with ODEs}, which are practically not influenced by the particular realization of the fractional derivatives or the integration time $\tau$ (at least up to $\tau=20$). However, as the fractional order of the dynamical system is slightly decreased (but remains close to the $\alpha=1$ value), we see that the phase space geometrical formations revealed by the Lagrangian descriptors become dependent on the integration time and the particular methods used to evaluate the fractional derivatives. To demonstrate how the phase space features change as we further decrease the fractional order $\alpha$,  we present in \cref{LDs for Duffing Oscillator with fractional differential equations of order 0.95 and 0.9} phase space plots similar to those seen in \cref{LDs for Duffing Oscillator with fractional differential equations of order close to one}, but for $\alpha=0.95$ [\cref{LDs for Duffing Oscillator with fractional differential equations of order 0.95 and 0.9 multiple}(a)--(d)] and $\alpha=0.9$ [\cref{LDs for Duffing Oscillator with fractional differential equations of order 0.95 and 0.9 multiple}(e)--(h)]. 

We note that in \cref{LDs for Duffing Oscillator with fractional differential equations of order 0.95 and 0.9 multiple}(a) and (e), where we present results obtained by the `nonlocal implicit inverse' method, we see some points colored in white. These points correspond to ICs for which we could not compute their backward-in-time evolution because the implemented root-finding method was not converging. Furthermore, we observe an increase in the number of these white points as we decrease the fractional order of the system from $\alpha=0.95$  [\cref{LDs for Duffing Oscillator with fractional differential equations of order 0.95 and 0.9}(a)] to $\alpha=0.9$  [\cref{LDs for Duffing Oscillator with fractional differential equations of order 0.95 and 0.9}(e)], which indicates that the `nonlocal implicit inverse' method has practically reached the limits of its applicability and cannot be efficiently applied to cases with smaller $\alpha$ values. Nevertheless, we note that the `time-reversing inverse' approach, used to create the results presented in the right three columns of \cref{LDs for Duffing Oscillator with fractional differential equations of order 0.95 and 0.9} can still be used for smaller $\alpha$ values. The phase space patterns depicted in \cref{LDs for Duffing Oscillator with fractional differential equations of order 0.95 and 0.9 multiple}(a) and (e) clearly show an increase in the disconnect of the branches that previously made the well-defined infinity-shaped structure, which becomes more pronounced for smaller values of $\alpha$, leading to patterns which are quite different from the ones seen for the integer-order case in \cref{LDs method for Duffing Oscillator with ODEs}. Nevertheless, the point symmetry of the created structures, which was observed in \cref{LDs for Duffing Oscillator with fractional differential equations of order close to one multiple}(a), (e), and (i), is still present.

The results obtained by the implementation of the Caputo derivative with the `time-reversing inverse' approach can be seen in \cref{LDs for Duffing Oscillator with fractional differential equations of order 0.95 and 0.9 multiple}(b)--(d) for $\alpha=0.95$ and in \cref{LDs for Duffing Oscillator with fractional differential equations of order 0.95 and 0.9 multiple}(f)--(h) for $\alpha=0.9$. We note that the phase portraits in all these figures retain the reflective symmetry with respect to the horizontal and vertical axes passing through the origin, as observed in the right three columns of \cref{LDs for Duffing Oscillator with fractional differential equations of order close to one}. For both $\alpha$ values, the infinity-shaped structure is still present and well-defined for $\tau=5$. [\cref{LDs for Duffing Oscillator with fractional differential equations of order 0.95 and 0.9 multiple}(b) and (f)]. For $\tau=10$ and $\alpha=0.95$ [\cref{LDs for Duffing Oscillator with fractional differential equations of order 0.95 and 0.9}(c)], we see the existence of an additional curve outside the infinity-shaped formation, similarly to what was observed in \cref{LDs for Duffing Oscillator with fractional differential equations of order 0.95 and 0.9 multiple}(g) and (k), but the distance between these two structures is larger than the ones observed in  \cref{LDs for Duffing Oscillator with fractional differential equations of order 0.95 and 0.9 multiple}(g) and (k). This additional curve is not present for $\tau=10$  when the order of the fractional derivative is reduced to $\alpha=0.9$ [\cref{LDs for Duffing Oscillator with fractional differential equations of order 0.95 and 0.9}(g)]. Furthermore, we hardly see any additional curves for $\tau=20$ both for  $\alpha=0.95$ [\cref{LDs for Duffing Oscillator with fractional differential equations of order 0.95 and 0.9}(d)] and $\alpha=0.9$ [\cref{LDs for Duffing Oscillator with fractional differential equations of order 0.95 and 0.9}(h)], in contrast to what was observed for the same integration time for $\tau=0.99$ [\cref{LDs for Duffing Oscillator with fractional differential equations of order close to one}(h)] and $\tau=0.98$ [\cref{LDs for Duffing Oscillator with fractional differential equations of order close to one}(l)].  The geometric formations in \cref{LDs for Duffing Oscillator with fractional differential equations of order 0.95 and 0.9 multiple}(f)--(h) 
show only infinity-shaped curves, which are similar to the patterns seen for the system with $\alpha=0.9999$ [\cref{LDs for Duffing Oscillator with fractional differential equations of order close to one multiple}(a)--(d)]. However, as the integration time increases [from \cref{LDs for Duffing Oscillator with fractional differential equations of order 0.95 and 0.9}(f) to \cref{LDs for Duffing Oscillator with fractional differential equations of order 0.95 and 0.9}(h)], we notice that the color of the curve becomes darker. The Lagrangian descriptors can still be computed for smaller values of $\alpha$ using the integration with the `time-reverse inverse' method but, as we have already mentioned, not with the `nonlocal implicit inverse' approach.

\subsection{Which method correctly reveals the system's dynamical behavior?}
\label{add}

Having computed the Lagrangian descriptors of the Duffing oscillator governed by fractional differential equations (\cref{fractional differential equation duffing oscillator}) using two different backward time integration methods, we see from \cref{LDs for Duffing Oscillator with fractional differential equations of order close to one} and \cref{LDs for Duffing Oscillator with fractional differential equations of order 0.95 and 0.9} that these methods produce qualitatively different results, where one yields broken NHIMs and the other still preserves their smoothness and continuity. The backward integration methods cause the differences. A natural question arises: Which method provides Lagrangian descriptors that correctly represent the phase space of the fractional-order differential equation dynamical system? 

\begin{figure}[tb!]
    \centering 
    \begin{subfigure}{1\textwidth}
        \includegraphics[width=0.95\textwidth]{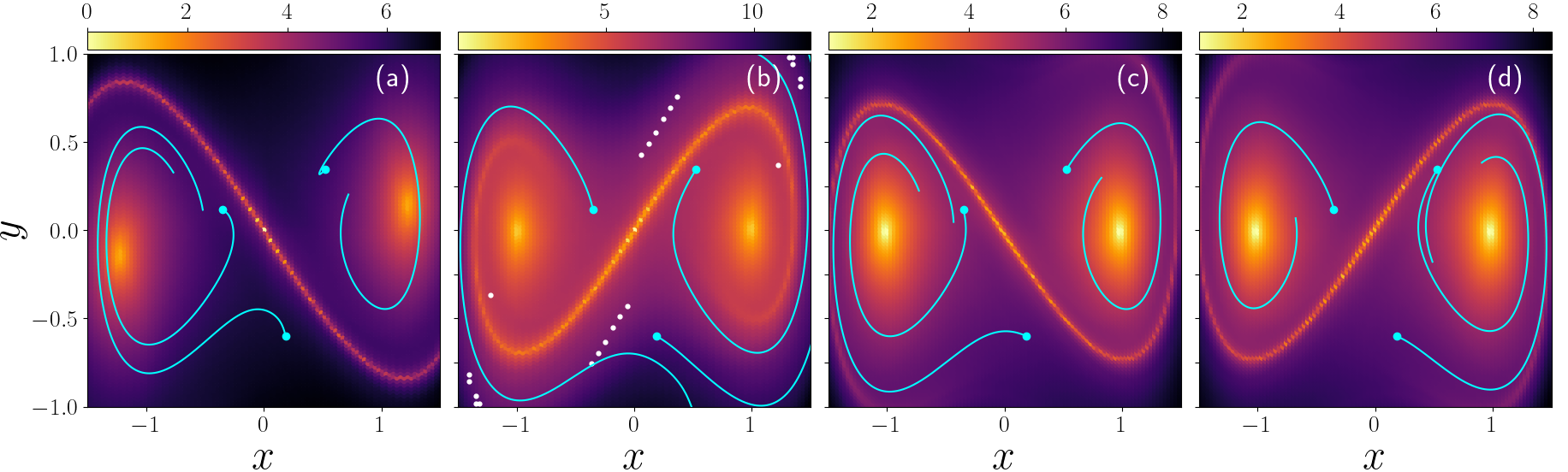}
    \end{subfigure}
    \caption{Initial conditions of the Duffing oscillator governed by the fractional differential equations \eqref{fractional differential equation duffing oscillator} with $\alpha=0.95$, colored according to their Lagrangian descriptor value (using the color scale above each panel). The plots are created using an equidistant grid of $100\times100$ points over the intervals $x\in[-1.5,1.5]$ and $y\in[-1,1]$, with an integration time of $\tau=5$. The Gr\"unwalled-Letnikov derivative \eqref{grunwald} is used for the forward time evolution of orbits to create a panel (a), and its 'nonlocal implicit inverse' version to obtain panel (b) for the backward time evolution of ICs. The results shown in panels (c) and (d) are obtained by the forward time implementation of the Caputo derivative \eqref{eq:Caputo} and the backward time application of its `time-reversing inverse' approach, respectively. We superimpose the forward [panels (a) and (c)] and backward [panels (b) and (d)] time evolution of three orbits (shown as cyan curves) with ICs $(x_0,y_0)=(-0.351,0.116)$, $(x_0,y_0)=(0.525,0.343)$, and $(x_0,y_0)=(0.186,-0.599)$ (indicated by cyan points) using the respective integration methods.}
    \label{Forward and backward LDs}
    \label[multifigures]{Forward and backward LDs multiple}
\end{figure}

To answer this question, we turn our attention to the Duffing oscillator with $\alpha=0.95$ and the phase portraits created using the Riemann-Liouville and Caputo fractional derivative with the `nonlocal implicit inverse' and the `time-reversing inverse integration' methods for $\tau=5$ in \cref{LDs for Duffing Oscillator with fractional differential equations of order 0.95 and 0.9 multiple}(a) and (b), respectively. For these cases, we analyze the relation of the phase space patterns created by the forward \eqref{forward p-norm LD definition}  and backward \eqref{backward p-norm LD definition} Lagrangian descriptors method, with respectively the forward and backward in time evolution of particular orbits. The obtained results are displayed in \cref{Forward and backward LDs}. 

The results of \cref{Forward and backward LDs multiple}(a) and (c) are obtained by the application of the forward Lagrangian descriptors method \eqref{forward p-norm LD definition} for the Gr\"unwald-Letnikov  \eqref{grunwald}  and Caputo  \eqref{eq:Caputo} derivatives, respectively. Furthermore, \cref{Forward and backward LDs multiple}(b) and (d) display the results for the backward Lagrangian descriptors method \eqref{backward p-norm LD definition} obtained using the `nonlocal implicit inverse' and the `time-reversing inverse' approaches, respectively. Three cyan dots and cyan curves corresponding to each dot are superimposed onto each Lagrangian descriptor scalar field. The curves indicate the time evolution of orbits associated with the ICs $(x_0,y_0)=(-0.351,0.116)$, $(x_0,y_0)=(0.525,0.343)$, and $(x_0,y_0)=(0.186,-0.599)$ (presented as cyan points).

The orbits that are evolved forward in time [shown in \cref{Forward and backward LDs multiple}(a) and (c) for both types of fractional derivatives] are attracted by the stable fixed points $(x,y)=(\pm1,0)$. Examining the backward-in-time evolution of these orbits in \cref{Forward and backward LDs multiple}(b), we observe consistency in their behavior, as these trajectories are now moving away from the fixed points $(x,y)=(\pm1,0)$. In particular, some of them, show a tendency  to approach the fixed point $(x,y)=(0,0)$ (which is a saddle in both the forward and the backward time evolution) before being pushed away by the origin's unstable manifold. In these three panels [\cref{Forward and backward LDs multiple}(a), (b), and (c)], the trajectories appear to be guided by structures revealed by the Lagrangian descriptors method.

On the other hand, we obtain inconsistencies in the observed orbital behaviors in \cref{Forward and backward LDs multiple}(d). Since the fixed points $(x,y)=(\pm1,0)$ are stable in the forward-in-time evolution [see \cref{Forward and backward LDs multiple}(c)], they must be repelling in the backward-in-time integration. However, we notice that in \cref{Forward and backward LDs multiple}(d), the cyan-colored orbits are also attracted by the fixed points $(x,y)=(\pm1,0)$. This inconsistent behavior informs us that the `time-reversing inverse' \eqref{time-reversing} 
only applies for integer values of $\alpha$. Hence, we can conclude that only the `nonlocal implicit inverse' method is the correct backward integration-in-time approach, yielding Lagrangian descriptors that reveal the phase space structures of dynamical systems governed by fractional differential equations.

Furthermore, we can now explain why the infinity-shaped NHIMs in \cref{LDs method for Duffing Oscillator with ODEs} for $\alpha=1$ are broken when $\alpha$ decreases, as shown in the first columns of \cref{LDs for Duffing Oscillator with fractional differential equations of order close to one multiple} and \cref{LDs for Duffing Oscillator with fractional differential equations of order 0.95 and 0.9 multiple}. 
The Duffing system with $\alpha=1$ (\cref{LDs method for Duffing Oscillator with ODEs}) has families of periodic orbits around the fixed points  $(x,y)=(\pm1,0)$ within the region bounded by the infinity-shaped manifolds. 
Introducing fractional derivatives 
destroys the closed-loop nature of the trajectories and leads to a stable spiral-looking dynamical evolution for $0<\alpha<1$. 
Consequently, the structures related to the connected stable and unstable manifolds for $\alpha=1$ become disconnected.

\section{Conclusions}
\label{C}

In this work, we studied the behavior of a dynamical system governed by fractional-order differential equations, namely the unforced and undamped Duffing oscillator \eqref{fractional differential equation duffing oscillator}, by applying, to the best of our knowledge for the first time to such systems, the method of Lagrangian descriptors. Using different definitions of fractional derivatives and approaches to perform the backward-in-time integration of orbits, we showed that the Lagrangian descriptors method can successfully reveal the underlying geometric features that govern the system's phase space transport. We also investigated in detail how the change of the order $\alpha$ of the used fractional derivatives, as well as the integration time $\tau$ needed for the computation of the Lagrangian descriptors, influence the morphology of the depicted geometrical structures. We observed that for values of $\alpha$ very close to the classical, integer-order Duffing oscillator ($\alpha=1$), we observe similar-looking phase space structures.  For values of $\alpha$ much smaller than $\alpha=1$, as well as when $\tau$ is increased,  we obtained phase space structures exhibiting large differences from the ones seen for the classical Duffing oscillator. We also showed that the `nonlocal implicit inverse' approach for integrating the fractional-order differential equations is the correct method to perform this task, although it showed some computational challenges that allow its implementation for very small $\alpha$ values. Finding efficient and accurate methods to perform the backward-in-time integration of fractional-order differential equations is an important task that we plan to address in the future, possibly by considering the right-handed fractional derivatives as an inverse of the corresponding left-handed ones \cite{frac1,wu2023right,wu2023terminal}. 

Our results clearly demonstrated the usefulness of the Lagrangian descriptors method in providing a qualitative geometric interpretation of the dynamics of dynamical systems governed by fractional-order differential equations, as it offers a way to examine phase space regions associated with different dynamical traits. We hope that our work will pave the way for more applications of the Lagrangian descriptors method to such systems. For example, we plan to extend the dynamical investigation of the fractional-order Duffing oscillator by considering $1<\alpha<2$ in a future publication.

\section*{Data availability}
The data that support the findings of this study are available
within the article and are available from the corresponding author
upon reasonable request.

\section*{Declaration of competing interest} 
The authors declare that they have no known competing financial interests or personal relationships that could have appeared to influence the work reported in this paper.

\section*{CRediT authorship contribution statement}
The manuscript was written with contributions from all authors. All authors have given their approval to the final version of the manuscript. \textbf{DT}: Software, Validation, Formal Analysis, Investigation, Writing - Original Draft; \textbf{HS}: Conceptualization, Methodology, Validation, Writing - Original Draft, Writing - Review \& Editing; \textbf{MA}: Conceptualization, Methodology, Validation, Writing - Original Draft, Writing - Review \& Editing; \textbf{CS}: Conceptualization, Methodology, Validation, Writing - Original Draft, Writing - Review.

\section*{Acknowledgements}

DT acknowledges the University of Cape Town for funding assistance. HS acknowledges support by Khalifa University through a Faculty Start-Up Grant (No.\ 8474000351/FSU-2021-011), a Competitive Internal Research Awards Grant (No.\ 8474000413/CIRA-2021-065), and a Research \& Innovation Grant (No.\ 8474000617/RIG-2023-031). He also acknowledges discussions with Ms.\ Sofwah Ahmad. MA acknowledges support  by the Society of Spanish Researchers in Southern Africa (ACE Sur de Africa), through the grant for FRA Visiting Lecturers Grants Program that they are sponsored by the Ramón Areces Foundation (FRA) and the Embassies of Spain in Angola, Mozambique, South Africa and Zimbabwe. 





\bibliography{references}

\end{document}